\documentstyle[aps,manuscript,epsfig]{revtex}
\begin{document}
\title{First Measurement of Antikaon Phase-Space 
Distributions  in Nucleus-Nucleus Collisions 
at Subthreshold Beam Energies}

\author{
M.~Menzel$^d$,
I.~B\"ottcher$^d$,
M.~D\c{e}bowski$^e$,
F.~Dohrmann$^{f,\S}$,
A.~F\"orster$^b$,
E.~Grosse$^{f,g}$,
P.~Koczo\'n$^a$,
B.~Kohlmeyer$^d$,
F.~Laue$^{a,*}$, 
L.~Naumann$^f$,
H.~Oeschler$^b$,
F.~P\"uhlhofer$^d$,
E.~Schwab$^a$,
P.~Senger$^a$,
Y.~Shin$^c$,
H.~Str\"obele$^c$,
C.~Sturm$^b$,
G.~Sur\'owka$^{a,e}$,
F.~Uhlig$^b$
A.~Wagner$^f$,
W.~Walu\'s$^e$\\
(KaoS Collaboration)\\ 
$^a$ Gesellschaft f\"ur Schwerionenforschung, D-64220 Darmstadt, Germany\\
$^b$ Technische Universit\"at Darmstadt, D-64289 Darmstadt, Germany\\
$^c$ Johann Wolfgang Goethe Universit\"at, D-60325 Frankfurt am Main, Germany\\
$^d$ Phillips Universit\"at, D-35037  Marburg, Germany\\
$^e$ Jagiellonian University, PL-30059 Krak\'ow, Poland\\
$^f$ Forschungszentrum Rossendorf, D-01314 Dresden, Germany \\
$^g$ Technische Universit\"at Dresden, D-01062 Dresden, Germany\\
$^{\S}$ Present address: Argonne National Laboratory, Argonne, IL 60439, USA\\
$^*$ Present address: Ohio State University, Columbus, OH 43210,   USA
}
\maketitle


\vspace{-1.cm}
\begin{abstract}
Differential production cross sections of 
K$^-$ and K$^+$ mesons have been measured 
as function of the polar emission angle
in Ni+Ni collisions  at a beam energy of 1.93 AGeV.
In near-central collisions, the spectral shapes and the widths of the 
rapidity distributions of K$^-$ and K$^+$ mesons
are in agreement with the assumption of isotropic emission.  
In non-central collisions,  the K$^-$ and K$^+$ rapidity distributions
are broader than expected for a single thermal source. In this
case, the polar angle distributions are strongly forward-backward peaked and 
the nonisotropic contribution to the total yield is about one third both for 
K$^+$ and K$^-$ mesons.  The K$^-$/K$^+$ ratio is found to be
about 0.03 independent of the centrality of the reaction. This value is  
significantly larger than predicted by microscopic transport calculations 
if in-medium modifications of K mesons are neglected.

\end{abstract}

\newpage

The behavior of strange particles in dense nuclear matter 
plays a crucial role in the dynamics of supernovae and for the 
stability of neutron stars \cite{glen_web_mos,bro_bet,li_lee_br,hei_jen}.
The creation of negatively charged kaons in
equilibrated neutron star matter becomes energetically possible if
the electro-chemical potential exceeds the effective
mass of kaons in the dense medium.
It was  speculated that a  
Bose condensation of K$^-$ mesons significantly  softens  
the equation-of-state of neutron star matter and hence catalyzes the
formation of low-mass black holes \cite{bro_bet}. This idea is based
on theoretical calculations predicting in-medium kaon-nucleon potentials  
which are repulsive for kaons but attractive for antikaons
\cite{brown1,schaffner,waas,lutz}.    

Experiments with  heavy-ion beams
provide the unique possibility to study the creation and propagation 
of strange particles in baryonic matter at densities well above 
nuclear matter density. In nucleus-nucleus collisions both the
total production cross sections and the angular distributions 
of strange mesons should be affected by the 
predicted in-medium potentials.
The sensitivity on in-medium effects 
is enhanced at beam energies  below  the kaon 
production threshold in nucleon-nucleon collisions 
(1.58 GeV and 2.5 GeV for the processes NN$\to$K$^+\Lambda$N and
NN$\to$K$^-$K$^+$NN, respectively).  
The creation of strange mesons is dominated by   
secondary processes (like pion-nucleon collisions) which happen more 
frequently at higher densities    
\cite{schaffner,li_ko_fang,cassing}.

In recent years, data on kaon and antikaon production in nucleus-nucleus 
collisions at subthreshold beam energies have been collected at the heavy-ion
synchrotron SIS at GSI in Darmstadt. The most remarkable findings are
that (i)  the K$^-$ yield is surprisingly high  
\cite{schroeter,barth,laue}, (ii) the in-plane flow of K$^+$ mesons is 
vanishingly small  
in Ni+Ni collisions at 1.93 AGeV
\cite{ritman}, (iii) preferential  out-of-plane emission of K$^+$ mesons 
is observed in Au+Au collisions at 1 AGeV \cite{shin} and (iv) 
the spectral slope of K$^-$ mesons is steeper than the slope of K$^+$ mesons 
in C+C collisions at 1.8 AGeV \cite{laue}. 
These observations  have all been explained as being due to modifications
of the K meson properties in the nuclear medium 
\cite{li_ko_fang,cassing,li_flow,li_ko_br,bra_ca_mo}.  

In this Letter we report on the measurement of 
K$^-$ and K$^+$ rapidity and angular  distributions 
in Ni+Ni  collisions at a beam energy of 1.93 AGeV. This energy is 
close to (below)  the production threshold for K$^+$ (K$^-$) mesons 
in nucleon-nucleon collisions. For the first time full rapidity distributions
of K$^+$ and K$^-$ mesons become available at SIS energies.
The production of K$^+$ mesons 
around target rapidity  in central Ni+Ni collisions at 1.93 AGeV  
was already studied  by the FOPI collaboration \cite{best}.  

The experiment was performed with the Kaon Spectrometer (KaoS) at the 
heavy-ion synchrotron at GSI 
in Darmstadt \cite{senger}. 
This magnetic spectrometer has a large acceptance
in solid angle and momentum ($\Omega\approx$30 msr, $p_{max}/p_{min}\approx$2).
Kaon decays in flight are minimized by the short distance of 5 - 6.5 m from 
target to focal plane. 
Separate measurements of velocity, momentum and time-of-flight provide 
particle identification and allow for an online rejection of events  with
only pions and protons in the spectrometer 
by factors of 10$^2$ and 10$^3$, respectively.
The background due to spurious tracks and pile-up is
strongly reduced by trajectory reconstruction based on
three large-area multi-wire chambers. The resulting 
signal-to-background ratio for K$^+$ and K$^-$ mesons is about 5:1 
and 10:1, respectively. The total efficiency of the trigger, of the track 
recognition and of detection amounts to about 75\%, both for K$^+$ and 
K$^-$ mesons. The loss of kaons decaying in flight is determined 
(and corrected) by Monte Carlo simulations.

The  $^{58}$Ni beam had an intensity of up to 3$\times10^8$ ions per spill
(8 seconds long) and impinged on a $^{58}$Ni target of 0.8 mm thickness.
The beam intensity was  similar for the K$^+$ and K$^-$  experiments and for
the different angle settings. The measurement was performed under four  
laboratory angles $\theta_{lab}$ = 32$^{\circ}$, 40$^{\circ}$, 
50$^{\circ}$ and  60$^{\circ}$
with 3 magnetic field settings  accepting a momentum range of 
280 MeV/c $< p_{lab} <$ 1120 MeV/c. The corresponding  
phase space covered by the experiment ranges in transverse momentum 
from about $p_t$ = 150 MeV/c to
700 MeV/c and in rapidity from about y$_{CM}$ = -0.65 to +0.2. The rapidity  
is defined as y$_{CM}$ = y$_{lab}$ - 0.5$\times$ y$_{proj}$ with
y$_{proj}$ the projectile rapidity (in the laboratory) and  
y$_{lab} = 0.5\times ln((E+p_z)/(E-p_z))$. 
Here, E is the  particle energy and 
$p_z$ is the particle momentum along the beam axis in the laboratory.
The value y$_{CM} = 0$  corresponds to midrapidity and 
y$_{CM} = \pm 0.89$ is target or projectile rapidity.

Approximatelly 60000  K$^+$ mesons and 8000 K$^-$ mesons have been registered. 
The large number of identified K mesons allowed the determination of  
differential cross sections in small intervalls of rapidity and 
transverse momentum.
The event ensemble is subdivided  in  near-central and
non-central reactions. The centrality of the collision is determined via
the multiplicity of charged particles measured in the interval
12$^{\circ} < \theta_{lab} < 48^{\circ}$ by a 
hodoscope consisting of 84 plastic-scintillator modules.
These particles - mostly protons and pions - have participated in the 
reaction and are therefore emitted to large angles.
From a measurement with a minimum bias trigger the 
reaction cross section has been determined for each bin of charged particle
multiplicity. As near-central collisions we  define the most
central 620$\pm$30 mb of the reaction cross section which corresponds
geometrically to  impact parameters smaller than 4.4 fm.

Figure 1  shows the cross sections for K$^+$ (upper panels) 
and K$^-$ production (lower panels) in near-central 
(left) and non-central (right) collisions as function of 
the transverse kinetic energy $m_T-m$ for five equally wide bins in rapidity 
as defined in the lower right panel.  The transverse mass is
given by $m_T = (p_T^2+m^2)^{(1/2)}$ with $p_T$ the transverse momentum
and $m$ the rest mass of the K mesons.
The spectral yields are scaled by factors of 10 (see upper left panel).
The error bars shown are due to statistics only. 
A systematic error of 10 \%
due to efficiency corrections and normalization procedures has to be added
to each data point. The lines represent 
Boltzmann distributions $d^3\sigma/dp^3 \propto exp(-m_T/T)$ fitted 
to the spectra individually for each rapidity bin above y$_{CM}$ = -0.54. 
The slope parameters of the spectra measured at -0.69 $<$ y$_{CM}$ $<$ -0.54 
are determined by linear extrapolation of the fitted $T$-values.     
The resulting values for $T$ are listed in Table 1.

In Figure 2 we show the multiplicity density distributions dN/dy of 
K$^+$ and K$^-$ mesons for 
near-central (left panel) and non-central (right panel) 
collisions as function of the rapidity y$_{CM}$. 
The particle multiplicity is given by N = $\sigma_K/\sigma_R$. 
The inclusive production cross section $\sigma_K$ is 
obtained by integration of the spectra presented in Figure 1 using 
the Boltzmann fits to  extrapolate to the unmeasured transverse mass regions.
The measured fraction of the multiplicity density varies with
rapidity. The values are given in Table 1. In total, about 70\% of the  
integrated multiplicity density distribution has been measured (taking  
advantage of the symmetry with respect to midrapidity).  
The error bars of the dN/dY distributions shown in figure 2 include systematic
uncertainties due to efficiency corrections and due to the extrapolation
procedure. The uncertainties due to beam normalization and acceptance 
corrections (which affect all data points) amount to 7\%  and have to 
be added. The error bars of the K$^-$/K$^+$ ratio do not include
contributions from systematic uncertainties which affect both 
K$^+$ and K$^-$ data.

For the reaction cross section  we use  $\sigma_R$=620 mb
for near-central collisions (as measured, see above) and
$\sigma_R$=$\sigma_{geo} - $620 mb for non-central collisions.   
The geometrical reaction cross section is defined as
$\sigma_{geo}$ = 4$\pi$(r$_0$ A$^{1/3}$)$^2$. 
For Ni+Ni collisions we use a radius parameter of r$_0$=1.24 fm
\cite{wagner} and obtain a value of $\sigma_{geo}$=2.9 b.
The inclusive cross sections for the production of K$^+$ and 
K$^-$ mesons in
Ni+Ni collisions at 1.93 AGeV  are $\sigma_{K^+}$ = 87$\pm$10 mb
and $\sigma_{K^-}$ = 2.8$\pm$0.4 mb corresponding to a mean multiplicity
of N$_{K^+} \approx$ 3$\times10^{-2}$ and 
N$_{K^-} \approx$ 1$\times10^{-3}$, respectively. The quoted errors include
all systematic uncertainties. 

The K$^+$ multiplicity density distribution for near-central
collisions shown in Figure 2 can be compared to K$^+$ data measured by the 
FOPI collaboration in central Ni+Ni collisions at 1.93 AGeV  
around target rapidity \cite{best}.  
The FOPI data were analyzed for the most central 350 mb of the 
cross section corresponding to impact parameters below 3.3 fm 
within a sharp cutoff model. The range of rapidities covered 
by the experiments FOPI and KaoS overlap only in a narrow region around
y$_{CM}$ = -0.6. Here the measured  K$^+$ multiplicity densities
agree with each other within the error bars which are about 28\% (FOPI) and 
12\% (KaoS) including systematic uncertainties. 

The solid lines in Figure 2 correspond to isotropic thermal distributions. 
The widths of the curves are given by the  inverse slope parameters of the  
spectra measured around midrapidity (see Figure 1).   
For near-central Ni+Ni collisions, the measured dN/dy distributions
of K$^+$ and K$^-$ mesons do not deviate significantly from
isotropy. For non-central collisions, however, the widths of the  
multiplicity density distributions are clearly broader than expected for an
isotropic thermal source both for  K$^+$ and K$^-$ mesons.

The lower panel in Figure 2 presents 
the K$^-$/K$^+$ multiplicity
ratio as function of the rapidity for near-central (left) 
and non-central (right) Ni+Ni 
collisions at 1.93 AGeV. The average  values for the K$^-$/K$^+$
ratios are  0.031$\pm$0.005 both for near-central and  non-central collisions.
Transport calculations performed by two 
different groups predict a value of about 0.01 for the K$^-$/K$^+$ ratio
in central Ni+Ni collisions at 1.93 AGeV when taking into account bare
(vacuum) masses for the K mesons \cite{li_brown,cass_brat}. 
These calculations 
include secondary production processes like $\pi$N$\to$K$^+$K$^-$N and  
strangeness exchange reactions $\pi\Lambda\rightleftharpoons$K$^-$N.
The measured K$^-$/K$^+$ ratios can be reproduced by microscopic transport
calculations  by choosing appropriate values for the strengths of the
repulsive K$^+$N potential and for the attractive K$^-$N potential
which are in principle free parameters in those models. 
Up to now, predictions either overestimate \cite{li_brown}  or underestimate 
\cite{cass_brat}  the measured K$^-$/K$^+$ ratio shown in Figure 2.

In a previous experiment we have studied the inclusive production of 
K$^+$ and K$^-$ mesons in C+C collisions at 2 AGeV \cite{laue}.
The reabsorption of K$^-$ mesons in this light system  
via the  strangeness exchange reaction K$^-$N$\to$Y$\pi$ 
(with Y = $\Lambda,\Sigma$) \cite{dover} should be much smaller than for 
the Ni+Ni system which is almost 5 times heavier 
(the K$^+$ mesons cannot be absorbed).  
Nevertheless, the observed values for the K$^-$/K$^+$ ratio  were found 
to be very similar  (K$^-$/K$^+$ = 0.038$\pm$0.018 for C+C at 2 AGeV). 
This observation indicates that the losses of K$^-$ mesons in the Ni+Ni system 
due to reabsorption are compensated by an enhanced production. 
It is interesting to note that in near-central 
Ni+Ni collisions the K$^-$/K$^+$ ratio is independent of transverse momentum
whereas in non-central Ni+Ni and in inclusive C+C collisions \cite{laue}
the  K$^-$/K$^+$ ratio decreases with increasing transverse momentum.

In order to quantify the deviation from isotropical K meson emission 
in the Ni+Ni system    we have calculated the  ratio
$R = \sigma_{inv}(E_{CM},\theta_{CM})$/$\sigma_{iso}(E_{CM})$
with $\sigma_{inv}$ the measured invariant kaon production cross section 
and $\sigma_{iso}$ a thermal isotropic distribution. 
The function $\sigma_{iso} \propto exp(-E_{CM}/T)$ is fitted to the data 
separately for  K$^+$ and K$^-$ and for near-central and non-central collisions.
In order to determine an polar emission pattern which is representative for
the bulk of the kaons, the values of $R$ are weighted with the kaon production
cross section at a given $E_{CM}$. Then, the $R$-values which fall into one 
bin of $\theta_{CM}$ are averaged. The resulting distributions  
$R(cos\theta_{CM})$ are shown in Figure 3 for K$^+$ (upper panel) and 
K$^-$ (lower panel) and for near-central (left) and non-central collisions 
(right). 

The dashed lines represent the function 
$R(cos\theta_{CM}) \propto 1 + a_2 \cdot \cos^2(\theta_{CM})$ which is 
fitted to the experimental distributions.  The anisotropy factors $a_2$ are 
listed in Table 2 together with the values for the anisotropic fraction of 
the total yield. The total yield is proportional to the  integral 
\begin{displaymath}
\int_{-1}^{+1} R(cos\theta_{CM}) dcos\theta_{CM} \propto (1 + a_2/3)
\end{displaymath}
and hence the nonisotropic contribution is calculated by ($a_2$/3)/(1+$a_2$/3).
The data in Figure 3 and the  result of their analysis show  that 
in near-central collisions the K$^-$
mesons were emitted isotropically whereas the emission of K$^+$ mesons
is slightly enhanced at forward-backward angles. In non-central collisions,
however,  the emission patterns of both K$^-$ and  K$^+$ mesons are   
strongly forward-backward peaked. In these cases the nonisotropic 
contributions contain approximately one third of the total yield.

The phase-space distributions of kaons and antikaons presented in this letter 
constitute a detailed dataset on strange meson production in heavy-ion 
collisions at SIS energies. When confronted with transport calculations, 
this dataset offers the possibility to test the model assumptions in order
to improve our understanding of the in-medium properties of hadrons.

In summary, we report the first measurement of
K$^-$ and K$^+$ meson multiplicities  as function of rapidity
and polar angle in heavy-ion collisions at SIS energies.
In near-central Ni+Ni collisions, the spectral shapes and the widths 
of the multiplicity density 
distributions of  K$^-$ and K$^+$ mesons are rather similar and in reasonable
agreement with the assumption of isotropic emission.
In non-central collisions, however, the K$^+$ and K$^-$ rapidity 
density distributions
are broader than expected for isotropic emission from a single thermal
source with a temperature given by the inverse slope parameter of the 
spectra measured around midrapidity.  
This effect is even more obvious in the polar angle distributions of
K$^+$ and K$^-$ mesons which are strongly forward-backward peaked
for non-central collisions. For both centrality selections, 
the K$^-$/K$^+$ ratio is about 0.03. This value is about three times larger than
predicted by transport calculations in which  in-medium modifications
of K$^+$ and K$^-$ mesons are neglected.

This work was supported by the German Federal Government (BMBF), by the
Polish Committee of Scientific Research (Contract No. 2P03B11515) and 
by the GSI fund for University collaborations.

\newpage       
\begin{table}
Table 1: Rapidity range y$_{CM}$, 
inverse slope parameters $T$  and measured fraction F$_{meas}$ 
of $K^+$ and K$^-$ mesons in near-central and non-central  
Ni+Ni collisions at 1.93 AGeV.
\begin{center}
\begin{tabular}{|c|c|c|c|c|c|}
         &near-central  & near-central & non-central  & non-central  &      \\
y$_{CM}$ &$T$(K$^+$) (MeV)&$T$(K$^-$) (MeV)&$T$(K$^+$) (MeV)&$T$(K$^-$) (MeV)&
F$_{meas}$\\
\hline
-0.69 $-$ -0.54 & 86$\pm$10 & 84$\pm$15 & 79$\pm$10 & 67$\pm$10 & $\approx$40\% \\
-0.54 $-$ -0.39 & 87$\pm$5  & 79$\pm$6  & 81$\pm$3  & 70$\pm$7  & $\approx$80\% \\
-0.39 $-$ -0.24 & 99$\pm$4  &100$\pm$6  & 89$\pm$4  & 73$\pm$7  & $\approx$80\% \\
-0.24 $-$ -0.09 & 96$\pm$4  &106$\pm$11 & 87$\pm$3  & 84$\pm$5  & $\approx$65\% \\
-0.09 $-$ +0.06 &107$\pm$7  & 94$\pm$12 & 97$\pm$6  & 79$\pm$6  & $\approx$40\% \\
\end{tabular}
\end{center}
\end{table}

\begin{table}
Table 2: Anisotropy factor $a_2$ and nonisotropic fraction of the total yield 
F$_{noniso}$ = ($a_2$/3)/(1+$a_2$/3) 
for K$^+$ and K$^-$ mesons in near-central and non-central Ni+Ni collisions 
at 1.93 AGeV.
\begin{center}
\begin{tabular}{|c|c|c|c|c|}
  & near-central & near-central & non-central & non-central \\
  & K$^+$     &     K$^-$    &    K$^+$    &   K$^-$    \\
\hline
$a_2$ & 0.6$\pm$0.2 & 0.2$\pm$0.2  & 1.6$\pm$0.2 & 1.7$\pm$0.4 \\
F$_{noniso}$ & 0.17$\pm$0.05 & 0.06$\pm$0.06 & 0.35$\pm$0.03 & 0.36$\pm$0.05 \\
\end{tabular}
\end{center}
\end{table}

\newpage
\vspace{1.cm}
Figure 1:
Production cross sections for K$^+$ (upper panel) and
K$^-$ mesons (lower panel) in near-central (left) and non-central (right) 
Ni + Ni collisions at 1.93 AGeV as function of the transverse kinetic
energy $m_T-m$. 
The data are sorted in bins of 
rapidity (as indicated in the lower right panel) 
and are scaled by factors of 10 (see upper left panel). 
The lines correspond to Boltzmann distributions fitted to the data (see text).

\vspace{1.cm}
Figure 2:
Multiplicity density distribution for K-mesons as function of  
rapidity in near-central (left)
and non-central (right) Ni+Ni collisions at 1.93 AGeV
(y$_{CM}$=0 corresponds to midrapidity).
Full data points are measured and  mirrored at y$_{CM}$=0 (open points).                                                     
Upper panel: K$^+$, 
middle panel K$^-$, lower panel: K$^-$/K$^+$ ratio.  
The K$^+$ and K$^-$ data for non-central collisions are multiplied 
by  a factor of 4. 
The lines correspond to isotropic thermal distributions with 
temperatures given by the inverse slope parameters of the spectra measured 
around midrapidity (T-values are indicated).

\vspace{1.cm}
Figure 3:
Polar angle distributions of $K^+$ (upper panel)
and $K^-$ mesons (lower panel) in the center-of-mass system
for near-central (left) and non-central (right) Ni+Ni collisions at 1.93 AGeV.
Full data points are measured and  mirrored at y$_{CM}$=0 (open points).
The lines correspond to the function
$R \propto 1 + a_2 \cdot \cos^2(\theta_{CM})$  fitted to the data (see text). 
The resulting values for $a_2$ are indicated.

\newpage
\begin{figure}[h]
\mbox{\epsfig{file=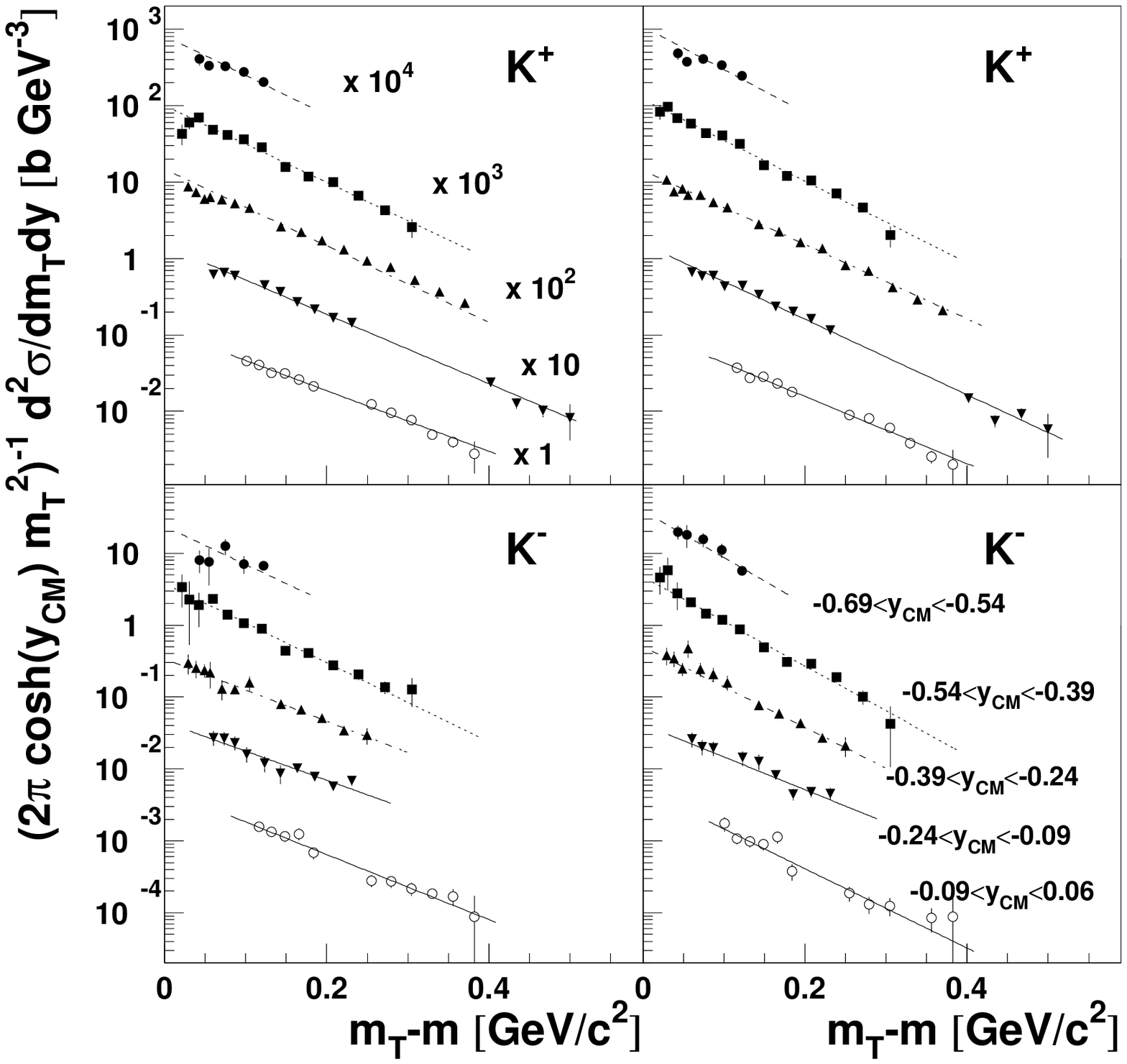,width=16.cm}}
\caption{}
\end{figure}

\begin{figure}[h]
\mbox{\epsfig{file=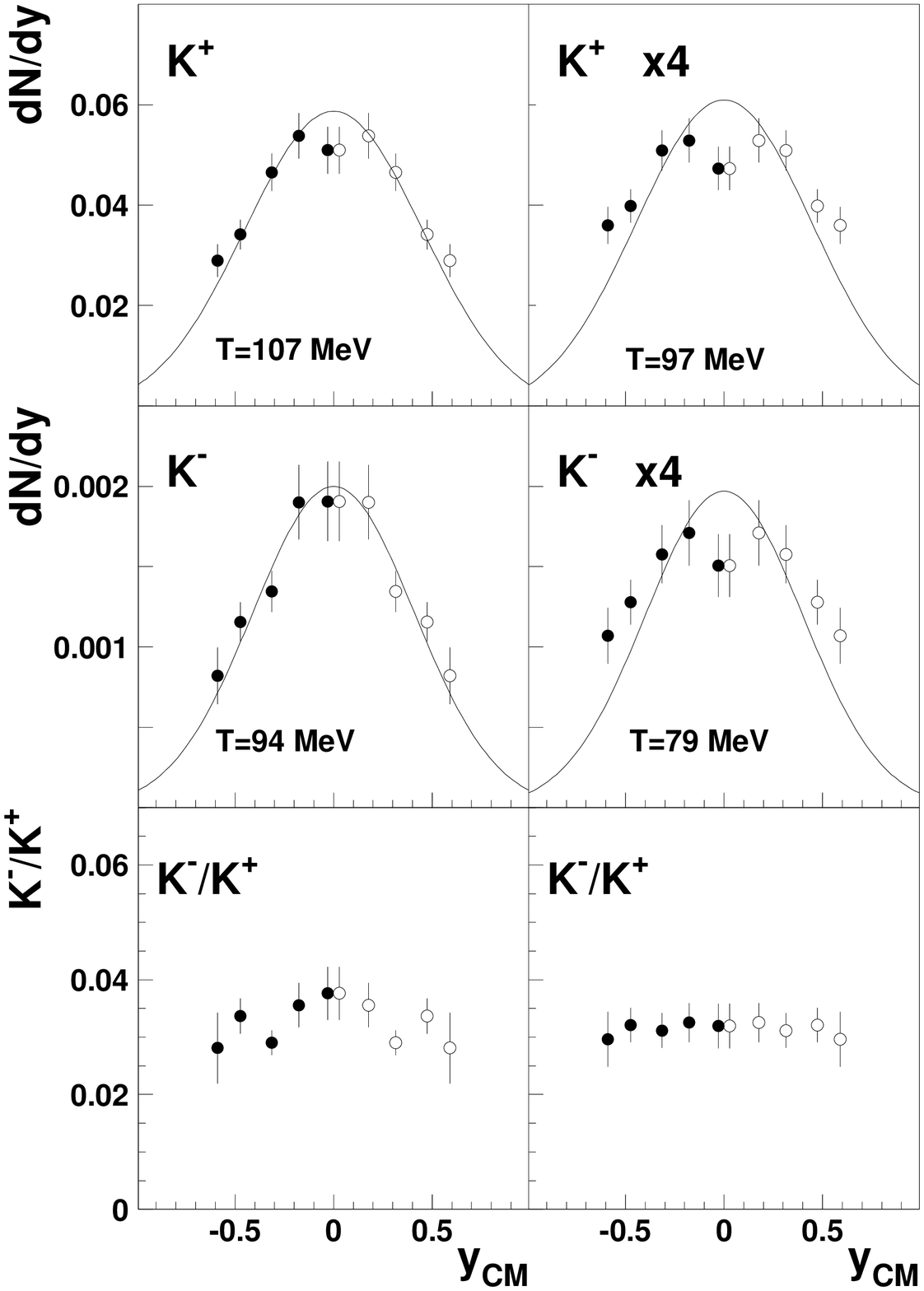,width=16.cm}}
\caption{}
\end{figure}

\begin{figure}[h]
\mbox{\epsfig{file=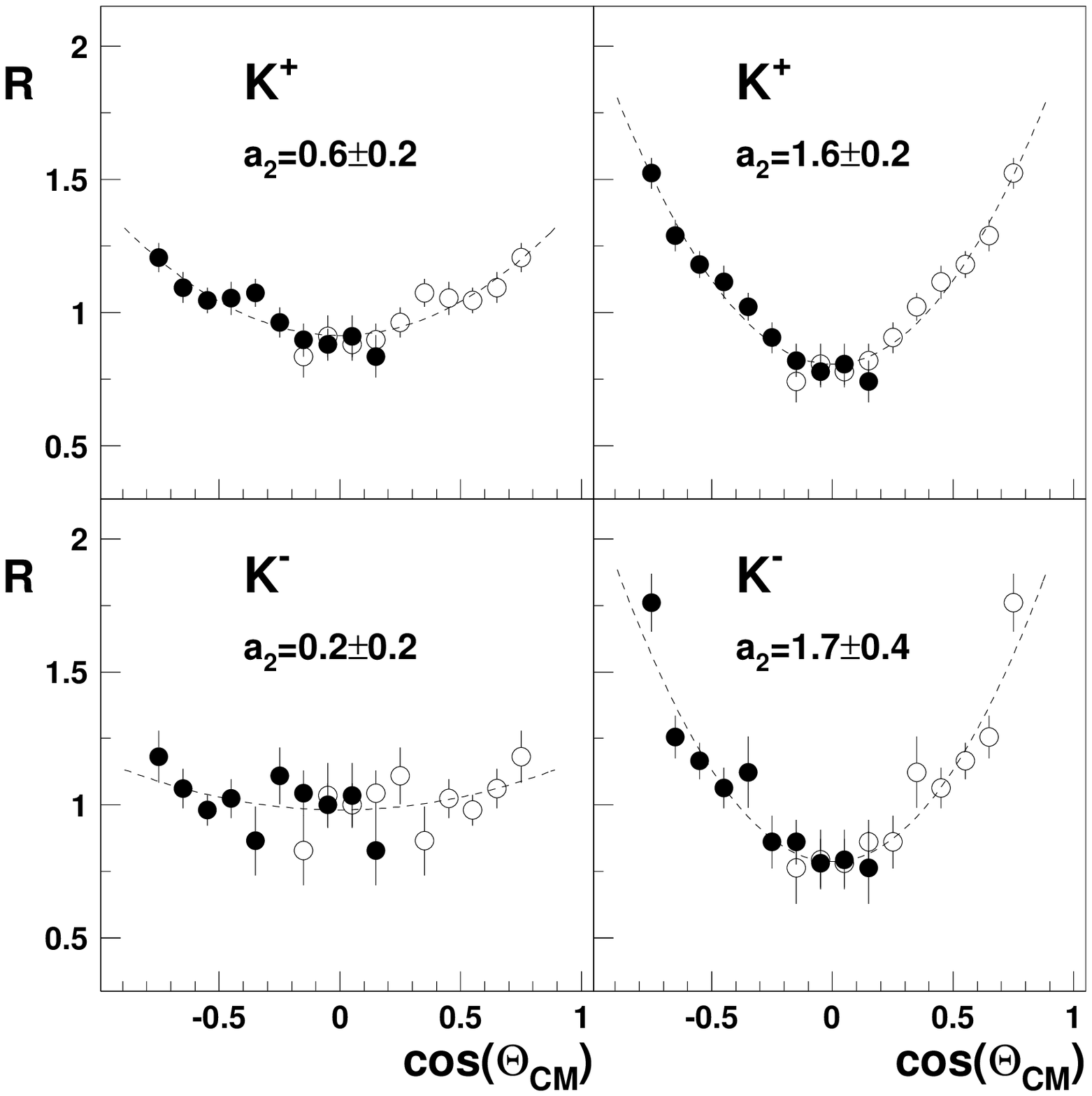,width=16.cm}}
\caption{}
\end{figure}

\end{document}